\documentclass[preprint]{aastex}			
\usepackage{psfig}

\received{1999 August 6}
\accepted{1999 October 4}

\slugcomment{Accepted for publication in Astrophysical Journal
Letters.\\ 1999 October 4.}

\shortauthors{Waddington et al.}
\shorttitle{A dusty radio source at $z=4.424$}

\newcommand{\vlaj}{\hbox{VLA~J123642$+$621331}}	
\newcommand{\etal}{et~al.}			
\newcommand{\muJy}{$\mu$Jy}			
\newcommand{\kmsmpc}{km s$^{-1}$ Mpc$^{-1}$}	
\newcommand{\hst}{{\it Hubble Space Telescope\/}} 

\begin{document}

\title{NICMOS Imaging of the Dusty Microjansky Radio Source \vlaj\ at
$z=4.424$\altaffilmark{1}}

\altaffiltext{1}{Based on observations with the NASA/ESA \hst\
obtained at the Space Telescope Science Institute, which is operated
by the Association of Universities for Research in Astronomy, Inc.,
under NASA contract NAS5-26555; and with the W. M. Keck Observatory,
which is operated as a scientific partnership among the University of
California, the California Institute of Technology, and NASA, made
possible by the generous financial support of the W. M. Keck
Foundation.}

\author{I. Waddington, R. A. Windhorst and S. H. Cohen}
\affil{Department of Physics and Astronomy, Arizona State University, 
PO Box 871504, Tempe, AZ~85287--1504}
\email{Ian.Waddington@asu.edu}

\author{R. B. Partridge}
\affil{Department of Astronomy, Haverford College, Haverford, PA~19041}

\and

\author{H. Spinrad and D. Stern}
\affil{Department of Astronomy, University of California, Berkeley, 
CA~94720}

\begin{abstract}

We present the discovery of a radio galaxy at a likely redshift of
$z=4.424$ in one of the flanking fields of the Hubble Deep Field.
Radio observations with the VLA and MERLIN centered on the HDF yielded
a complete sample of microjansky radio sources, of which about 20\%
have no optical counterpart to ${\rm I}\le 25$ mag.  In this Letter,
we address the possible nature of one of these sources, through deep
\hst\ NICMOS images in the F110W (J$_{110}$) and F160W (H$_{160}$)
filters.  \vlaj\ has a single emission line at 6595~\AA, which we
identify with Ly$\alpha$ at $z=4.424$.  We argue that this faint (${\rm
H_{160}}=23.9$~mag), compact ($r_e\simeq$ 0\farcs2), red (${\rm I_{814}} -
{\rm K} = 2.0$) object is most likely a dusty, star-forming galaxy with
an embedded active nucleus.

\end{abstract}

\keywords{galaxies: active --- galaxies: starburst --- galaxies:
evolution --- galaxies: individual (VLA~J123642$+$621331)}

\section{Introduction}

One of the contemporary topics of interest in extragalactic astronomy
is to measure the star formation history of the universe
\citep{Madau96}.  However, a major uncertainty in calculating the star
formation rate (SFR) is the effect of dust obscuration, particularly
for measurements in the ultraviolet and at high redshifts.  For
example, \citet{Calzetti99} found that at $z\sim3$ even a modest
amount of dust can reduce the 1500~\AA\ flux of a galaxy by a factor
of 5.  This is of particular significance for the high-$z$ evolution
of the SFR: does the SFR turn over at $z\sim1.5$ as \citet{Madau96}
suggested, or is it constant (or even increasing) for $z\ga1.5$ (e.g.\
Pascarelle, Lanzetta, \& Fern\'{a}ndez-Soto 1998; Steidel \etal\
1999)?\nocite{Pascarelle98,Steidel99} If a high-redshift starburst is
{\it significantly\/} reddened by dust, then its ultraviolet flux will
be so obscured that it will have a negligible Lyman-limit break. Such
sources would not have been detected by \citet{Steidel99}, and their
measurements of the SFR could still be under-estimating the actual
amount of star formation at high-$z$.

A powerful technique to avoid the problem of dust obscuration is to
use radio-selected galaxies as probes of the star formation history
\citep{Cram98}.  The most important advantage of this technique is
that radio emission is not attenuated by dust and thus a
radio-selected sample is unbiased with respect to dust, unlike the
optical/infrared selection techniques that are widely used.
\citet{Haarsma99} conclude from their study that the radio-selected
SFR is somewhat higher than the \citet{Madau96} results for $z<1$,
even when the latter have been corrected for dust.  At higher
redshifts, the ``radio Madau diagram'' is poorly constrained but is
consistent with a relatively constant or increasing star formation
rate.  This suggests that a significant fraction of the star formation
in the universe may be obscured by dust.

Radio-selected samples thus provide an essential tool for
understanding the cosmic history of dust, as well as of the SFR.  In
particular, radio sources that are heavily reddened in the optical
(restframe ultraviolet at redshifts of interest) can be used to
constrain the dust content of high-$z$ galaxies.  We present in this
Letter deep {\it HST}/NICMOS images of the microjansky radio source
\vlaj.  At a probable redshift of 4.424, this is the second highest
redshift radio-selected galaxy currently known.  It was barely
detected in the reddest optical wavebands and is an excellent
candidate for a dusty high-$z$ galaxy.

We use a cosmology with ${\rm H}_0=65$~\kmsmpc, $\Omega_{\rm M}=0.2$
and $\Omega_{\rm \Lambda}=0$ throughout.  All magnitudes are in the AB
system, unless noted otherwise.  For comparison, Vega-based magnitudes
are approximately given by: ${\rm I_{814}}-0.4$, ${\rm J_{110}}-0.7$,
${\rm H_{160}}-1.3$, ${\rm K}-1.9$, and $({\rm I_{814}}-{\rm K})+1.5$.

\section{\vlaj} 

Two of the deepest high-frequency radio surveys available are those
centered on the Hubble Deep Field \citep{Richards98,Richards00} and
the Small Selected Area 13 \citep{Windhorst95}.  These surveys reach
5$\sigma$ detection limits of 8~\muJy\ at 8.5~GHz (HDF and SSA13) and
40~\muJy\ at 1.4~GHz (HDF).  Optical identifications of the sources
were made on deep \hst\ and ground-based images, primarily in the
I-band.
Approximately 20\% of the 1.4~GHz sources were not identified,
although a few of them show faint optical emission below the formal
completeness limit of the I-band images.  The full sample is discussed
by \citet{Richards99a}; in this Letter we investigate in detail one of
these ``unidentified'' sources.

VLA J123642$+$621331 is a steep-spectrum radio source
($\alpha=0.94\pm0.06$) with flux densities of 70~\muJy\ at 8.5~GHz and
470~\muJy\ at 1.4~GHz \citep{Richards98,Richards00}.  Approximately
10\% of its radio flux density is extended in an eastward jet in
0\farcs15 MERLIN observations, with the remainder in a resolved
compact core \citep{Muxlow99}.  The source was identified with a faint
object below the formal completeness limit of the HDF Flanking Field
I-band image (IW3), with a magnitude of ${\rm I}_{814}=25.3\pm0.2$
(Figure~\ref{image}a).  The object was also identified in the K-band
image of \citet{Dickinson98} from the KPNO 4-meter telescope, and has
a magnitude of ${\rm K}=23.25\pm0.05$ (Figure~\ref{image}d).  The KPNO
images do not detect the object in either the J or H bands.
\citet{Aussel99} identify \vlaj\ in their supplementary list of
sources detected by ISOCAM on the Infrared Space Observatory.  It has
a flux of 23$^{+10}_{-12}$~\muJy\ in the LW3 filter at 15~\micron,
corresponding to an AB magnitude of $20.5\pm0.5$.  The source lies at
the edge of the SCUBA sub-millimeter map of \citet{Hughes98} and was
not detected, giving a conservative upper limit of 5~mJy to its
850~\micron\ flux.

\section{NICMOS Observations \& Processing}

In December 1997, we observed \vlaj\ with the \hst\ NICMOS camera 2
for three orbits in F110W (close to the J-band) and six in F160W
(essentially the H-band), in the Continuous Viewing Zone.  In each
orbit, five 1024-second exposures were taken using a spiral dither
pattern with 1\farcs3 offsets.  We reduced the images using a two
stage process: first we ran them through the standard pipeline
CALNICA, using the most recent calibration files, and then we applied
a further flat-field correction to the data, adapted from standard
ground-based infrared imaging methods.

Following CALNICA processing, additional bad pixels were flagged after
a visual inspection of the data --- these consisted of a dead column,
the coronograph hole and other insensitive pixels (also known as the
``grot'').  Since the NICMOS camera actually consists of four
physically separate sub-arrays, we obtained better results by dividing
each partially-reduced image into four separate quadrants.  For each
filter, all the exposures of each quadrant were stacked and a median
image calculated.  Given that the exposures were dithered by 1\farcs3,
this produced a map of the residual instrumental features, devoid of
any astronomical objects --- a ``super-sky''.  The super-sky images
for each quadrant were normalized to the mean of all four quadrants in
order to preserve the quadrant-to-quadrant photometric accuracy.  Each
exposure was then {\it divided\/} by this normalized super-sky.  We
also tested the results of {\it subtracting\/} a scaled copy of the
super-sky from each exposure and found that the results were not
significantly different (the standard deviation of an empty sky region
varied randomly by less than 3\% between the two methods).  The
success of this super-sky division suggests that the instrumental
features left after CALNICA processing were most likely due to
differences between the sky and the calibration flat-fields, rather
than poor subtraction of the bias or dark current.  Finally, all the
quadrants were combined to produce F110W and F160W mosaics.  By
ensuring that the mean of our super-sky was unity, the overall
sensitivity of the images was unchanged by the additional processing
and they were calibrated using the most recent photometric calibration
parameters from STScI.

Figure~\ref{image}c shows the 6-orbit NICMOS image in F160W.  The
counterpart to the radio source is clearly detected at 1.6~\micron,
with an AB magnitude of ${\rm H}_{160}= 23.87\pm 0.04$ in a 1.5 arcsec
diameter circular aperture.  The 3-orbit NICMOS image in F110W
provided a marginal detection of ${\rm J}_{110}=25.2\pm0.4$
(Figure~\ref{image}b).  Thus the object is very red with $({\rm
I}_{814}-{\rm K})=2.0\pm0.2$, $({\rm J}_{110}-{\rm
H}_{160})=1.3\pm0.4$ and $({\rm H}_{160}-{\rm K})=0.6\pm0.1$.

\section{Keck Observations}

We obtained spectra of \vlaj\ through 1\farcs5 wide,
$\approx$~30\arcsec\ long slits using the LRIS spectrograph
\citep{Oke95} at the Keck~II telescope in slitmask mode.  On UT 1998
February 19, with 0\farcs8 seeing and photometric conditions, we
observed \vlaj\ for 1.9~hr (position angle $103\degr$) with the 400
lines mm$^{-1}$ grating, sampling the wavelength range 5700--9400~\AA,
at $\Delta\lambda_{\rm FWHM} \approx 11$~\AA.  On UT 1999 May 10, with
0\farcs6 seeing and thin cirrus, we observed the source for 2~hr
(position angle $-67\degr$) with the 150 lines mm$^{-1}$ grating, with
$\Delta\lambda_{\rm FWHM} \approx 17$~\AA\ resolution over the
wavelength range 4000~\AA\ to 1~\micron.  Between each 1800~s
exposure, we performed a $\sim4\arcsec$ spatial shift along the slit
to facilitate removal of fringing in the reddest regions of the
spectra.  Final wavelength calibration is accurate to better than
1~\AA.

\begin{figure}
\psfig{file=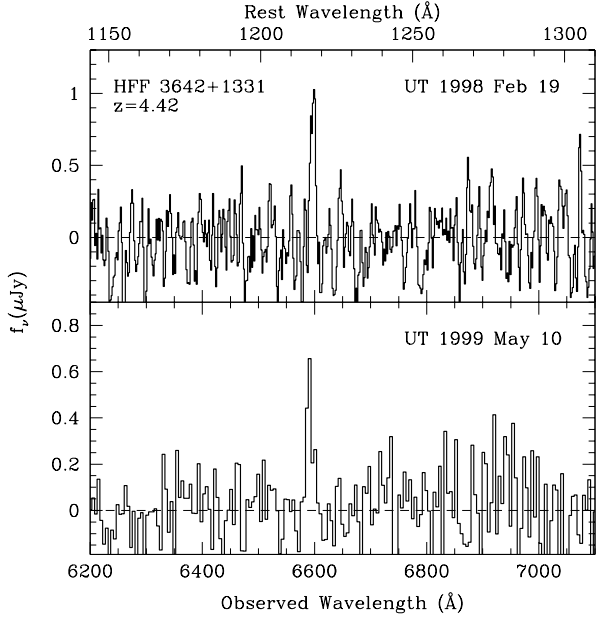,width=88mm}
\parbox[b]{88mm}{\caption{Keck spectra of \vlaj.  The single emission
line is detected in two independent observations over a period of more
than a year, and is identified as Ly$\alpha$ at $z = 4.424$.  The 1998
spectrum is smoothed with a 3-pixel boxcar filter. \label{spectrum}}}
\end{figure}

There is a strong, single emission line at $\lambda \simeq 6595$~\AA\
{\it in both data sets,} which we identify with Ly$\alpha$ at a
redshift of 4.424 (Figure~\ref{spectrum}, Table~\ref{table}).  The line
shifts by $\approx 7$~\AA\ between the two observations, which
corresponds to a velocity difference of 320~km~s$^{-1}$ if it is due
to a Doppler shift.  In both years' data, the line was offset by
$\approx 1\arcsec$ to the north-west of the (marginal) I-band
detection.  Emission-line regions of high-redshift radio galaxies are
known to be kinematically complex \citep{Chambers90,vanOjik97}, thus
slight pointing changes between the two observations may have caused
the slit to sample different regions of spatially--extended,
line-emitting gas.  

\section{Discussion}

The only two reasonable identifications for the single emission line
are Ly$\alpha$ or [\ion{O}{2}] 3727.  We argue that it is unlikely to
be [\ion{O}{2}] at $z=0.77$ for the following reasons.  The faint K
magnitude of 23.3 argues strongly against the source being at low-$z$
when it is compared to the K--$z$ relation of radio galaxies (see van
Breugel \etal\ 1999\nocite{vanBreugel99} for a recent version).  If
\vlaj\ were at $z=0.77$, it would be 3--4 magnitudes underluminous in
K compared with all other known radio galaxies at that redshift.  The
restframe equivalent width if the line were [\ion{O}{2}] would be
$W^{\rm rest}_{\rm [O II]} > 207$~\AA, which is very large for
[\ion{O}{2}].  The absence of a redshifted [\ion{O}{3}] doublet at
$\sim$8860~\AA\ further argues against $z=0.77$, although this
argument is not completely satisfying as galaxies show a wide range in
[\ion{O}{3}]/[\ion{O}{2}] ratios and our flux limits are not
particularly strong.

\begin{figure}
\psfig{file=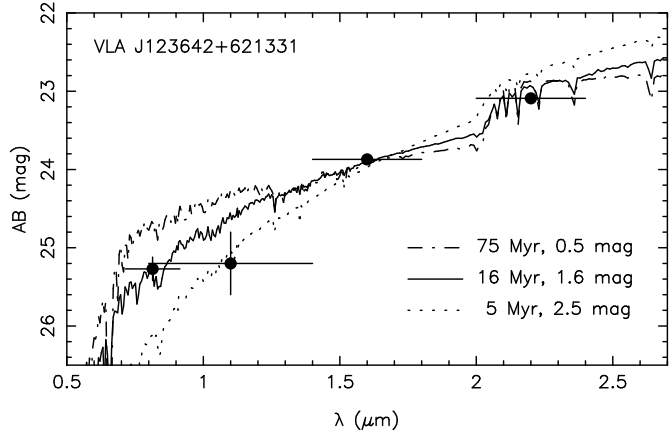,width=88mm}
\parbox[b]{88mm}{\caption{Optical/infrared spectral energy
distribution of the radio galaxy.  The three spectra are the
best-fitting model (solid line), the oldest \& least dusty model
(dot-dash line) and youngest \& most dusty model (dotted line), both
at the 3-$\sigma$ confidence limits. \label{sed}}}
\end{figure}

The alternative is to identify the emission line with Ly$\alpha$ at
$z=4.424$.  We compared the observed SED from 0.8~\micron\ to
15~\micron\ with the 1996 revision of the spectral evolution models of
\citet{Bruzual93}.  We used a single burst model of solar metalicity
and added a foreground screen of dust, modeling the effects of dust
obscuration with the extinction law of Calzetti, Kinney, \&
Storchi-Bergmann (1994)\nocite{Calzetti94}.  The best-fitting model we
obtain has an age of $(1.6\pm1.0)\times10^7$ years with
$A(V)=1.6\pm0.3$ mag (Figure~\ref{sed}).  No model of any age is able
to reproduce the red colors of this galaxy with $A(V)<0.5$~mag at the
99.99\% confidence level.  The results are essentially independent of
the metalicity, from 0.02 of solar to solar.  We note that the models
predict a 15~\micron\ flux that is too faint, although given the large
ISO errors the models were consistent to within $\la2\sigma$.  We also
attempted to fit the galaxy with a model spectrum at $z=0.77$ and were
unable to obtain {\it any\/} acceptable fit to the
optical/near-infrared SED: the minimum $\chi^2$ was $\sim30$ for a
18~Gyr model with no dust extinction, compared with a $\chi^2$ of 2.5
for the best-fitting $z=4.424$ model.  In particular, the low redshift
models could not reproduce the red ${\rm H}_{160}-{\rm K}$ color of
$0.6\pm0.1$, nor did they fit the 15~\micron\ flux.

The sub-millimeter (850~\micron) to radio (1.4~GHz) spectral index has
recently been used by \citet{Carilli99} as a redshift indicator. With
a spectral index of $\alpha^{353}_{1.4}\ge-0.4$
($S_\nu\propto\nu^{-\alpha}$), \vlaj\ is consistent with this
correlation at either $z=0.77$ for a starburst galaxy, or at $z=4.424$
if the 1.4~GHz flux is dominated by an AGN.  Thus the current SCUBA
sub-millimeter detection limit does not favor either a high or low
redshift interpretation.  The lack of a far-infrared/sub-millimeter
detection prevents us from drawing conclusions about the emission
properties of the dust that must be present based on the reddening of
the optical/infrared SED.

We note that it is possible that we detected a serendipitous $z=4.42$
background object in the Keck spectra, that is not associated with the
radio source.  We used the number counts and redshift distribution of
galaxies in the HDF to find the random probability that a $z>4$ source
with a magnitude 1--2~mag below the completeness limit of the I-band
image would be found within a 1\arcsec\ radius of the optical
identification.  From this calculation we estimate that there is a
probability of $\la10^{-3}$ that the spectra are of a background
source.

\begin{figure}
\psfig{file=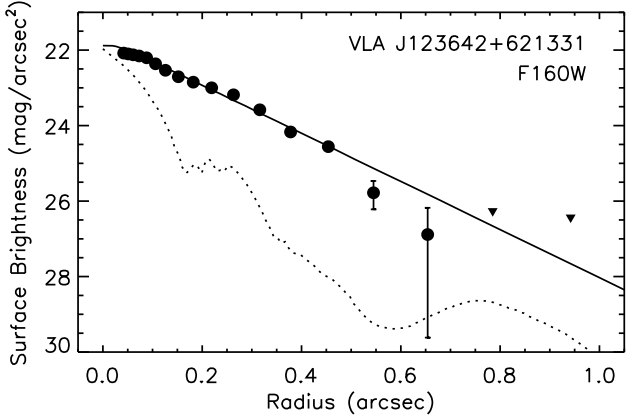,width=88mm}
\parbox[b]{88mm}{\caption{The NICMOS F160W surface brightness profile
of \vlaj.  Triangles denote 1-$\sigma$ upper limits.  The best-fitting
model (solid line) is an exponential with scale-length 0\farcs17
(1.4~kpc), convolved with the appropriate PSF (dotted line) from
``Tiny Tim''. \label{profile}}}
\end{figure}

The surface brightness profile of \vlaj\ was produced by fitting
elliptical isophotes to the F160W image (Figure~\ref{profile}).  We
extracted model surface brightness profiles from an ensemble of
two-dimensional exponential and de~Vaucouleurs models convolved with a
model PSF from ``Tiny Tim''.  The best-fitting model is an exponential
with scale-length $r_e=0\farcs17\pm0\farcs02$ (where $I\propto
e^{-r/r_e}$) or $1.4\pm0.2$~kpc in our chosen cosmology.  Any
unresolved point source (i.e., an AGN) contributes $\la 10$\% of the
total flux in the H-band.  A de~Vaucouleurs $r^{1/4}$ profile does not
fit the data.  We note briefly the contrast between the dominant
exponential profile of \vlaj\ and the profiles of other very red
high-$z$ sources, of which several are now known at $z\simeq1.5$--2
that have regular $r^{1/4}$ profiles (e.g., Stiavelli \etal\ 1999,
Waddington \etal\ 1999\nocite{Stiavelli99,Waddington99}).

\begin{figure}
\psfig{file=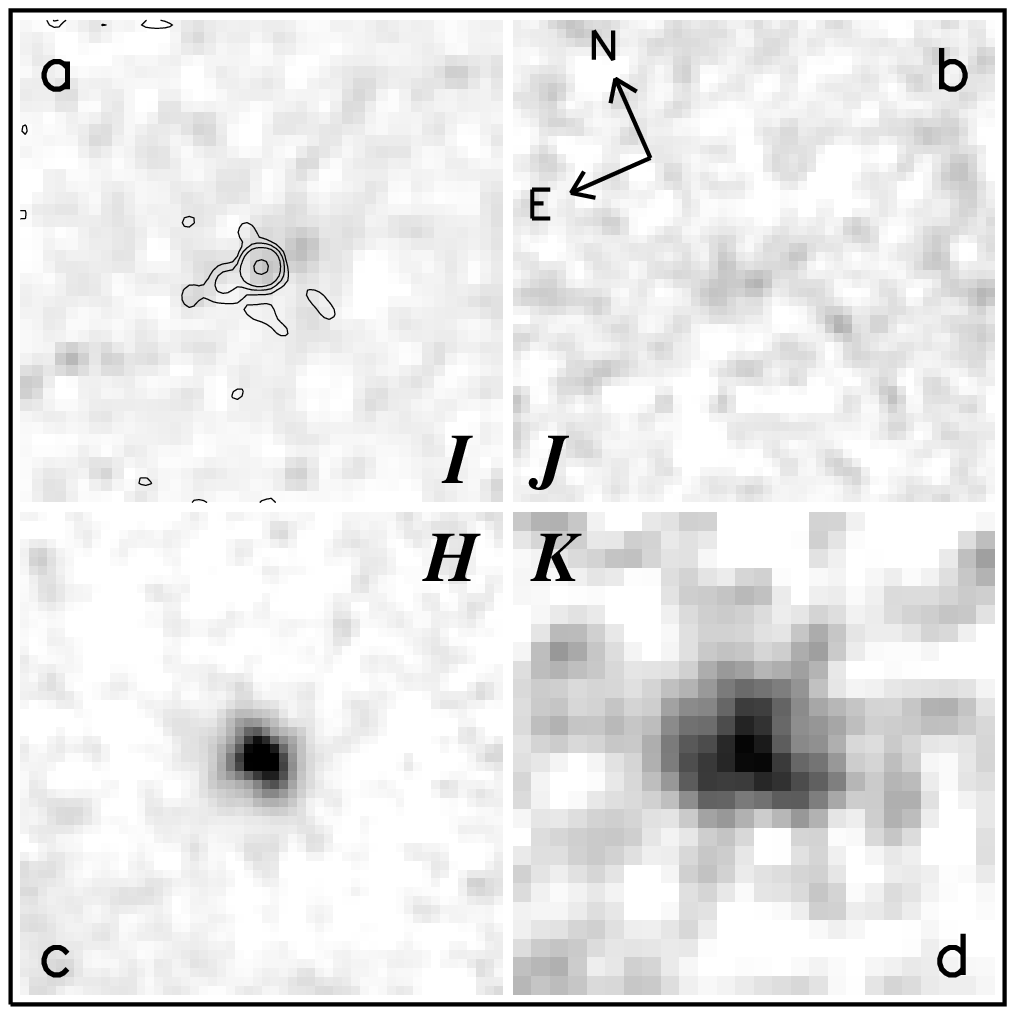,width=88mm}
\parbox[b]{88mm}{\caption{Multiband images of \vlaj.  (a) WFPC2 F814W
(I-band) grey-scale image, overlaid with the 1.4~GHz combined (uniform
weighted) MERLIN/VLA radio image of \citet{Muxlow99}, at a resolution
of approximately 0.15~arcsec. Contour levels are at flux densities of
8, 16, 32, 128~\muJy\ ($\sigma=4$~\muJy).  (b) The source is barely
detected in a 3-orbit NICMOS F110W (J-band) image.  (c) The 6-orbit
NICMOS F160W (H-band) image shows a very red object at the radio
position.  (d) Ground-based K-band image with the KPNO 4-meter
(1\farcs0 seeing). Each image is 4~arcsec on a side and has been
smoothed with a 2-pixel wide gaussian. \label{image}}}
\end{figure}

It can be seen in Figure~\ref{image} that there is some
two-dimensional structure to the galaxy.  It is extended in a roughly
north--south direction in the F160W image, perpendicular to the radio
jet.  In the F814W image there is a faint hint of extended emission to
the west of the radio position, in the opposite direction to the radio
jet but aligned with the Ly$\alpha$ emission.  The marginal J-band
detection may be similarly extended to the west.  The apparent
structure in the K-band is probably not real, but is likely due to the
PSF and/or the drizzling procedure, given that other objects in the
image show a similar structure.  It is also possible that there is a
contribution to the K-band light from redshifted [\ion{O}{2}] 3727.
Most high-$z$ radio galaxies have distorted or aligned morphologies
(e.g., Best, Longair, \& R\"{o}ttgering 1997\nocite{Best97}) and it is
an important question whether such a low power radio source as \vlaj\
is also aligned with its ultraviolet/optical emission.

The preceding arguments lead us to the following interpretation.
\vlaj\ is a disk system at a likely redshift of $z=4.424$, containing
an embedded, weak AGN.  With an 8.4~GHz luminosity of
$2.0\times10^{25}$~W~Hz$^{-1}$, it is at the faint end of the radio
luminosity function for AGN, intermediate between an AGN and a
starburst.  Both the radio jet that is visible in the MERLIN image and
the sub-millimeter to radio spectral index suggest it is an AGN.  The
AGN must be obscured in the ultraviolet/optical, as it is not seen in
either the F160W profile or in the colors of the SED.  The observed
Ly$\alpha$ luminosity of $2\times10^{42}$~erg~s$^{-1}$
(Table~\ref{table}) is a factor of 10--100 fainter than typical
high-redshift radio galaxies \citep{Rottgering97}, and its Ly$\alpha$
line-width (${\rm FWHM}\simeq440$~km~s$^{-1}$) is 2--3 times narrower
than the other known $z>4$ radio galaxies \citep{vanBreugel99}.

The galaxy underwent a burst of star formation approximately 16
million years before we observe it, although the presence of dust
suggests that this was not the first starburst.  It is a dusty galaxy
with an extinction of 1.6~mag in the visual, and a corresponding
ultraviolet extinction of $A(1216~$\AA$)\simeq4$~mag.  This would be
sufficient to render any Ly$\alpha$ emission {\it from the disk\/}
undetectable in our Keck spectrum, if it were of comparable intrinsic
luminosity to the (unreddened) line that we do observe.  The
Ly$\alpha$ emission line came from an extended region about 1\arcsec\
to the north-west of the galaxy, roughly aligned with the radio axis.
This Ly$\alpha$ emission region could be an infalling gas cloud that
is either scattering/reradiating Ly$\alpha$ from the AGN into our line
of sight, or is the site of recent (jet-induced) star formation.  Such
extended Ly$\alpha$ clouds have been observed around several
high-redshift AGN at $z\simeq 2.4$ (Francis, Woodgate \& Danks 1997;
Windhorst, Keel, \& Pascarelle 1998; Keel \etal\
1999)\nocite{Francis97,Windhorst98,Keel99} and we should perhaps
expect such structures to be more abundant at higher redshifts, where
the gas has had less time to collapse.  Deep imaging of the source at
$\sim6600$~\AA\ is needed in order to investigate the nature of this
cloud.

{\acknowledgments We thank Eric Richards, Ken Kellerman, Ed Fomalont
and Tom Muxlow for useful discussions and for sharing unpublished
data; Andy Bunker for contributing to the Keck observations; and Mark
Dickinson for making his KPNO infrared observations of the HDF
publicly available.  This work was supported by NASA grant
GO-7452.0*.96A from STScI under NASA contract NAS5-26555, and by NSF
grant AST9802963.}




\begin{deluxetable}{lcc}
\tablewidth{125mm}
\tablecaption{Spectroscopic Measurements}
\scriptsize \tablehead{ \colhead{Parameter} & \colhead{UT 1998 Feb 19}
& \colhead{UT 1999 May 10}} 

\startdata 

$z$ \dotfill & $4.427 \pm 0.001$ & $4.421 \pm 0.001$ \\ 
$\lambda$ (\AA) \dotfill & $6597.4 \pm 1.0$ & $6590.4 \pm 1.1$ \\
$F_{{\rm Ly}\alpha}$ (10$^{-17}$ erg cm$^{-2}$ s$^{-1}$) \dotfill &
$0.74 \pm 0.14$ & $0.48 \pm 0.11$ \\
$F_\lambda^{\rm cont}$ (10$^{-21}$ erg cm$^{-2}$ s$^{-1}$ \AA$^{-1}$)
\dotfill & $-0.2 \pm 7.3$ & $8.4 \pm 4.8$ \\
$W_{{\rm Ly}\alpha}^{\rm obs}$ (\AA) \dotfill & $> 483$ & $>249$ \\
FWHM$_{{\rm Ly}\alpha}$ (km s$^{-1}$) \dotfill & $420 \pm 75$ & $464
\pm 155$ \\
$L_{{\rm Ly}\alpha}$ (10$^{42}$ erg s$^{-1}$) \dotfill & $2.4 \pm 0.4$
& $1.5 \pm 0.4$ \\

\enddata
\label{table}
\end{deluxetable}

\end{document}